\documentclass[11pt]{iopart}
\usepackage{iopams}
\usepackage{epsfig}
\newcommand{\be}{\begin{equation}}
\newcommand{\ee}{\end{equation}}

\newcommand{\BE}{\begin{eqnarray}}
\newcommand{\EE}{\end{eqnarray}}

\newcommand{\erf}{{\rm erf}}
\newcommand{\id}{{\rm 1\!\!I}}

\newcommand{\bx}{\ensuremath{\mathbf{x}}}

\newcommand{\avg}[1]{\left\langle{#1}\right\rangle}
\newcommand{\davg}[1]{\left\langle\left\langle{#1}\right\rangle\right\rangle}

\begin{document}
\title[Statistical mechanics of a model eco-system]{Statistical mechanics and stability of a model eco-system}

\author{Yoshimi Yoshino\dag, Tobias Galla\ddag~ and Kei Tokita\dag\S
}

\address{\dag Graduate School of Science and Cybermedia Center, Osaka University, Toyonaka, Osaka 560-0043, Japan\\ 
\ddag\ The Abdus Salam International Center for 
Theoretical Physics, Strada Costiera 11, 34014 Trieste, Italy\\ 
\S\ Graduate School of Frontier Biosciences,  
Osaka University, Suita, Osaka 565-0871, Japan}

\begin{abstract}
We study a model ecosystem by means of dynamical techniques from
disordered systems theory. The model describes a set of species
subject to competitive interactions through a background of resources,
which they feed upon. Additionally direct competitive or co-operative
interaction between species may occur through a random coupling
matrix. We compute the order parameters of the system in a fixed point
regime, and identify the onset of instability and compute the phase
diagram. We focus on the effects of variability of resources, direct
interaction between species, co-operation pressure and dilution on the
stability and the diversity of the ecosystem. It is shown that
resources can be exploited optimally only in absence of co-operation
pressure or direct interaction between species.
\end{abstract}


\ead{\tt yoshimi@cp.cmc.osaka-u.ac.jp, galla@ictp.it, tokita@cmc.osaka-u.ac.jp}

\section{Introduction}
Models of interacting individuals can be understood as many-body
systems of statistical mechanics, and tools developed originally in
the context of physics may be employed to address their dynamics and
stationary states. This approach has fruitfully been applied to a
variety of agent-based models inspired by economics and game theory,
see e.g. the recent textbooks \cite{Book1,Book2,Book3}. Attention here
focuses on the interplay of co-operation and competition between
interacting agents, and on the efficiency of their use of external
information and resources. Statistical mechanics here offers a variety
of valuable tools to study the global co-operative behaviour of such
systems, and to understand their phase structure. In particular
disordered systems theory \cite{Book4} allows one to address interacting
agent-models in which interaction matrices are drawn from random
ensembles, and to compute typical average quantities for such
models. Real-world systems are of course not random, but highly
correlated. The aim of statistical mechanics approaches is hence often
not to study specific instances, but rather the general properties of
classes of models as a function of the parameters characterising the
distribution from which couplings are drawn. One may ask for example
whether quantities such as connectivity, homogeneity or the strength
of interaction affect the stability of a given model system. Taking
an ensemble average here in a sense corresponds to studying all
possible realisations of a given model at the same time, and hence to
making statements about effects of model parameters in general, as
opposed to analyses of specific real-world instances.

This approach has been used to study e.g. the effects of
self-interaction and memory in models of financial trading
\cite{Book1,Book2} or to examine how co-operation pressure, order of
interactions impacts on the stability and trajectories of replicator
systems of evolutionary game theory \cite{DO,OD,OD2}. In the context
of population dynamics models with random interactions were first
addressed by May in \cite{May}.

In this paper we study a model of a simple food-web composed of
species and resources, originally proposed in a more basic form in
\cite{andreamatteo}. 
The level of the resource consumption by species and its relationship with the stability of 
the ecosystem and the species richness is one of the main issues in ecology \cite{Begon}. In \cite{andreamatteo}, interaction between species is not through direct interaction (e.g. via prey-predator relations) but exclusively through
the use and dependence on resources. If for example species $A$
consumes a resource which $B$ feeds upon as well, then this introduces
a negative and symmetric interaction between $A$ and $B$. The strength
of negative interaction between $A$ and $B$ is hence regulated by the
overlap in their dependence of resources.

Due to the symmetry of interactions the discussion of
\cite{andreamatteo} focuses on a static analysis of this model
eco-system. Here we choose a complementary dynamical approach, which
allows us to address a broader class of interaction modes.  Static
studies necessarily rely on the existence of a Lyapunov function,
extremised by the trajectories of the ecosystem, and are hence limited
to systems with symmetric interactions. In the case of an ecosystem
this is an obvious drawback, as competitive interaction of
e.g. prey-predator pairs can not appropriately be addressed. A direct
study of the dynamical equations allows us to extend the analysis to
cases of asymmetric interaction matrices, and in particular to discuss
the effects of anti-correlation on the behaviour of the
system \cite{galla,gallahebb}. Asymmetric interaction come in two ways in the present
eco-system. Firstly, we introduce direct interaction between species,
in addition to the indirect interaction through the use and dependence
on resources. Secondly, we study the effects of possibly asymmetric
dilution of the network of interacting species.

The aim of our work is here twofold. Firstly, the study of the present
model extends the statistical mechanics analysis of existing
replicator models \cite{DO,OD,OD2,galla,gallahebb}, and relates to
studies of Minority Games \cite{Book1,Book2}. Complex phase behaviour
and different patterns of ergodicity breaking and instabilities have
been identified in such models, with similarities as well as
differences between replicator-type models and other systems. One
purpose of the present work is thus to contribute to the
classification of such models according to the different types of
phase transitions they exhibit, and to identify possible universal
features.

Secondly, the model system here has a clear-cut ecological
interpretation, even though the model may be criticised for not
accurately capturing many features of real-world eco-networks. While
our approach is a dynamical one, and ultimately results in a
stochastic process for a single `effective' species, all disorder in
our model is quenched, i.e. the interaction web and coupling strengths
are fixed at the beginning, and replicator dynamics are then considered
on this fixed network. This approach on the one hand makes the model
analytically tractable and allows one to reduce the description to a
set of a few non-linear equations describing the relevant order
parameters. One the other hand it constitutes a considerable
restriction with respect to real-world eco-system, in which the web of
interaction is of course not fixed, but subject to dynamical evolution
itself, requiring the study of the dynamics of the network itself in
combination with population dynamics on the network. Such evolving
food-web models have for example been presented in
\cite{Caldarelli,McKane2000,sole2002,McKane2004,Drossel2004,McKane_Drossel_2005}. Related work is also found in \cite{Manrubia1,Manrubia2,Jain}. Results here rely mostly on numerical simulations (see however \cite{McKane2000} for descriptions in a Master equation formalism) and the
food-webs resulting from these models have been compared to real-world
data with respect to quantities such as the number of trophic levels,
their relative populations and the typical connectivity of species.
These models, some of which combine initial Gaussian random score
matrices with evolving species networks, clarified the necessary
conditions of types of functional responses and dietary choices
(specialist/generalist) for producing realistic webs, whose structure
agreed with empirical data.  

From the technical point of view it is interesting to note
that recent stochastic models of complex food-webs
\cite{McKane2000,sole2002} and the `neutral' model \cite{Volkov2003}
effectively reduce multispecies stochastic process to a `one species'
process of a representative species which is subject to a `mean-field'
interaction with the remaining system, and that these models derive
reasonable species abundance distributions in good agreement with real
data.  In a similar fashion our approach reduces the dynamics of
species randomly coupled via quenched interaction to a 'one species'
effective process as well. This mapping is fully exact in the
thermodynamic limit in the statistical sense. Apart from providing a
starting point for more realistic modifications of the present model,
our analysis can hence, to a certain degree, be seen as complementary
to the approach of \cite{McKane2000,sole2002}.

The paper is organised as follows: we will first define the model, and
then briefly discuss the statistical mechanics analysis based on a
path-integral approach. We then turn to a stability analysis, and then
discuss the effects of resource variation, direct interaction between
species, co-operation pressure and dilution in the subsequent
sections. We summarise our results in the conclusions section and
point out lines for potential future research.

\section{Model Definitions}
The model describes an eco-system consisting of $N$ species, labelled
by $i=1,\dots,N$ and $P=\alpha N$ resources $\mu=1,\dots,\alpha
N$. $\alpha$ is here a model parameter and is taken not to scale with
$N$, i.e. we assume $\alpha={\cal O}(N^0)$. The composition of the
eco-system at time $t$ is described by concentrations $x_i(t)$ of
species $i=1,\dots,N$, which evolve in time according to the following
replicator equations \cite{Book5}
\be\label{eq:repl}
\frac{\dot x_i(t)}{x_i(t)}=f_i[{\bf x(t),Q(t)}]+\nu(t).
\ee
$f_i$ here denotes the fitness of species $i$ at time $t$, and is
frequency dependent. To be more precise $f_i$ is taken to depend on
the composition of the ecosystem ${\bf x(t)}=(x_1(t),\dots,x_N(t))$ as
well as on the abundance of resources ${\bf
Q}(t)=(Q^1(t),\dots,Q^P(t))$. $\nu(t)$ is a global `field' variable,
which is (up to a sign) typically chosen as the mean fitness in order to maintain the
overall concentration of species.

\begin{figure}[t!!!]
  \vspace*{10mm} ~~~~~~~~~~~~~~~~\epsfxsize=60mm
  \epsffile{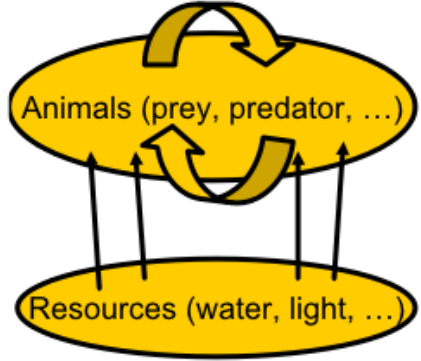} \vspace*{2mm}
  \caption{Illustration of the model: species compete for scarce resources while at the same time being subject to direct interaction e.g. through prey-predator relations.}
\label{fig:model}
\end{figure}

We will in the following assume that the fitness of species is
composed of three contributions $f_i(t)=f_{i,s}[{\bf
x(t)}]+f_{i,r}[{\bf Q(t)}]+f_{i,c}(x_i(t))$. $f_{i,s}$ denotes a term
describing direct species interaction, $f_{i,r}$ refers to
interaction due to competition for resources. These two components of
the model are illustrated in Fig. \ref{fig:model}, and can be understood similar to what is referred to as basal and intermediate species for example in \cite{McKane2004}. $\alpha$ thus controls the relative number of basal species (resources) over intermediate species in our model. Finally $f_{i,c}$
is an additional contribution describing an external so-called
cooperation pressure, driving the eco-system to a state in which all
species are present at equal concentration.  We will in the following
detail these three contributions to the fitness further.

Following \cite{OD} we will choose the direct interaction between species to be characterised by a random couplings, i.e.
\be
f_{i,s}[{\bf x}]=\sum_{j=1}^N w_{ij}x_j,
\ee
where the matrix elements $\{w_{ij}\}$ are chosen from Gaussian ensembles according to the following distribution
\be
P(w_{ij},w_{ji})=\frac{N}{2\pi \sqrt{w^2(1-\Gamma^2)}}\exp\left(-\frac{N(w_{ij}^2-2\Gamma w_{ij}w_{ji}+w_{ji}^2)}{2w^2(1-\Gamma^2)}\right)
\ee
for any pair $i<j$. The diagonal elements are taken to vanish,
$w_{ii}=0$. Denoting the average over the random couplings by an
overbar $\overline{\cdots}$ one thus has
\be
\overline{w_{ij}}=0, ~~~ \overline{w_{ij}^2}=\frac{w^2}{N}, ~~~ \overline{w_{ij}w_{ji}}=\Gamma\frac{w^2}{N}.
\ee
$\Gamma$ is a symmetry parameter and takes values $\Gamma\in [-1,1]$. For $\Gamma=1$ the interaction between any pair of species $i<j$ is fully symmetric, $w_{ij}=w_{ji}$. For $\Gamma=0$ $w_{ij}$ and $w_{ji}$ are uncorrelated, and $\Gamma=-1$ corresponds to a prey-predator relation, $w_{ij}=-w_{ji}$. Choosing intermediate values of $\Gamma$ allows one to interpolate smoothly between these regimes. The ecologically most relevant setup presumably corresponds to negative values of $\Gamma$, describing competitive direct interaction between species, rather than co-operation.

The second contribution $f_{i,r}$ to the fitness of species $i$ describes its propensity to reproduce due to the presence or otherwise of resources. We here follow the lines of \cite{andreamatteo}. Let us assume that the amount by which species $i\in\{1,\dots,N\}$ relies on resource $\mu\in\{1,\dots,P\}$ is described by a coefficient $\xi_i^\mu$, with large $\xi_i^\mu$ signalling a strong dependence of $i$ on $\mu$. Then we will take $f_{i,r}[{\bf Q}]$ to be of the form
\be
f_{i,r}[{\bf Q}]=\frac{1}{N}\sum_\mu \xi_i^\mu Q^\mu(t).
\ee
In turn a large abundance of $i$ will then deplete the abundance of $\mu$ so that we write
\be
Q^\mu(t)=Q^\mu[{\bf x(t)}]=Q_0^\mu-\sum_j\xi_j^\mu x_j(t).
\ee
$Q_0^\mu$ here denotes the abundance of resource $\mu$ in absence of
species and the second term on the right-hand side corresponds to the
consumption of resource $\mu$ by the different species
$j=1,\dots,N$. Recall that large $\xi_j^\mu$ indicates that species
$j$ consumes resource $\mu$ at a high rate, thus a large concentration
$x_j(t)$ (equivalently, a large number of individuals of species $j$)
adds to the depletion of resource $\mu$. The availability $Q^\mu(t)$
of resource $\mu$ thus becomes time-dependent, as the concentrations
of species $\{x_j(t)\}$ evolve in time. In particular it appears
interesting to ask the question whether or not the system is able to
organise in a state which avoids over- and under-exploitation of
resources, i.e. a state in which all $Q^\mu(t)$ remain close to zero
asymptotically. We will address this question below. Following our
earlier approach we take the coefficients $\{\xi_i^\mu\}$ to be drawn
from a random distribution, specifically we choose them to be
independent Gaussian variables, with mean $q$ and unit variance, i.e.
\be
\overline{\xi_i^\mu}=q, ~~~~ \overline{(\xi_i^\mu)^2}-(\overline{\xi_i^\mu})^2=1.
\ee
According to the above remarks they describe the interaction
between the species layer of the eco-system and the resource
layer. While the following analysis focuses mostly on the case of
Gaussian $\{\xi_i^\mu\}$ the generating functional theory below and
computer simulations show that only the first two moments of the
$\{\xi_i^\mu\}$ are relevant, so that more general distributions can
be addressed as well with the methods used here. To complete the
definition of $f_{i,r}$, it remains to specify the
$\{Q_0^\mu\}$. Following
\cite{andreamatteo} we write
\be\label{eq:defqnull}
Q_0^\mu=P+\sigma\sqrt{P}\zeta^\mu
\ee
with $\{\zeta^\mu\}_{\mu=1,\dots,P}$ independent standard Gaussian variables. The model parameter $\sigma$ thus controls the variability of resources. The scaling with $P=\alpha N$ of the $\{Q_0^\mu\}$ is chosen to guarantee a well defined thermodynamic limit, with which the theoretical analysis will eventually be concerned.

Finally, we will study the effects of co-operation pressure on the eco-system. This variables acts to suppress the growth of individual species and is incorporated by a contribution
\be
f_{i,c}(x_i)=-2ux_i
\ee
to the fitness of species $i$ \cite{OD,Book6}. In an ecological
setting $u$ takes mostly positive values denoting intra-species
competition (but see also a comment on potential settings with
negative $u$ below). For $u\to\infty$ the ecosystem is found in a
state of perfect co-operation and maximal diversity (with all species
surviving and having equal concentrations). As we will confirm later,
a reduction of $u$ leads to a smaller number of surviving species, and
hence a reduced diversity. In order to obtain a complete overview of
the phase behaviour of the model, we extend the analysis to negative
values of $u$ as well.

The definition of the dynamics (\ref{eq:repl}) is completed by stating the choice of $\nu(t)$ we will make in the following. In the analysis of the statics of the model it was found that only states with the normalisation $N^{-1}\sum_i x_i=\alpha/q$ contribute to the thermodynamics of the system \cite{andreamatteo}. Accordingly, we also restrict the dynamics to such configurations, and choose initial conditions and the subsequent Lagrange parameters $\{\nu(t)\}_{t\geq 0}$ such that the constraint
\be\label{eq:norm}
\frac{1}{N}\sum_{i} x_i(t)=\frac{\alpha}{q}
\ee
is fulfilled at all times. This amounts to the choice $\nu(t)=-\frac{q}{\alpha}\frac{1}{N}\sum_i x_i(t) f_i[\bx(t),{\bf Q}(t)]$ .

To conclude the presentation of the model let us briefly point out
some of its obvious limitations. Firstly, due to the Gaussian choices
of the $\{\xi_i^\mu\}$ and of the $\{Q_0^\mu\}$ negative values of
these quantities might statistically occur (in the cases of the
abundances $\{Q_0^\mu\}$ this is however suppressed in the
thermodynamic limit due to the scaling with $N$ in
(\ref{eq:defqnull})). Secondly, the replicator dynamics
(\ref{eq:repl}) do not guarantee that all $Q^\mu(t)$ remain positive
at all times. These drawbacks are consequences of the solubility of
the model, as models with non-Gaussian disorder at the same scaling
with $N$ or additional constraints on the resource abundances are
difficult to treat analytically. We would however like to note that
with our choice of parameters (e.g. $q=1$) most of the $\{\xi_i^\mu\}$
are indeed positive. The model is furthermore invariant under
simultaneous shifts of the means of all $\{Q_0^\mu,\xi_i^\mu\}$ so
that their averages can be chosen sufficiently high as to minimise the
amount of negative couplings.

\section{Generating functional and effective species process}\label{sec:gfa}
\subsection{Effective macroscopic theory and fixed point ansatz}
The model lends itself nicely to the study by the tools of disordered systems theory. For fully symmetric couplings $\Gamma=1$ one identifies 
\be
{\cal H}=\frac{1}{2N}\sum_\mu Q^\mu (t)^2 - \frac{1}{2}\sum_{i\neq j}
w_{ij} x_i(t) x_j(t)+u\sum_i x_i(t)^2 
\ee
as a Lyapunov function, minimised by the replicator dynamics (\ref{eq:repl}).  Thus the stationary state of the model can in this case be obtained by purely static considerations based on replica theory. For general symmetry $\Gamma$ no such Lyapunov function can be found, and the analysis needs to deal directly with the microscopic dynamics. The method of choice is here based on generating functionals, originally proposed in the context of random replicators in \cite{OD}, and recently used in \cite{galla,gallahebb}. The analysis focuses on the dynamic partition function
\be
Z[{\bf \Psi}]=\davg{\exp\left(i\int dt \sum_i x_i(t)\Psi_i(t)\right)}
\ee
where the average $\davg{\cdot}$ extends over all trajectories of the
system permitted by the equations of motion. ${\bf \Psi}$ is a source
field introduced to generate dynamical correlation functions, and
$Z[{\bf \Psi}]$ is hence the Fourier transform of the probability
measure on the space of paths generated by the replicator
equations. $Z[{\bf \Psi}]$ can then efficiently
be averaged over the disorder, and evaluated by the method of steepest
descents in the thermodynamic limit $N\to\infty$. We will not enter
the detailed mathematics here, but will only report the final
outcome\footnote{Imposing the above normalisation (\ref{eq:norm}) ensures that no super-extensive terms are found in the generating functional analysis and that the usual saddle-point integration can be carried out in the thermodynamic limit.}. One finds a description in terms of effective single-species
trajectories, described by the following multiplicative Gaussian
stochastic process
\be
\dot x(t)=x(t)\bigg(\int_{t_0}^t dt' R(t,t') x(t') -\eta(t)+\nu(t)\bigg)\label{eq:effproc}.
\ee
($t_0$ denotes the time at which the dynamics is started). The key components are the retarded interaction kernel
\be\label{eq:kernel}
R(t,t')=-2u\delta(t-t')-\Gamma w^2 G(t,t')-\alpha(\id-G)^{-1}(t,t')
\ee
and the coloured Gaussian noise $\{\eta(t)\}$ which exhibits temporal correlations of the form
\be
\avg{\eta(t)\eta(t')}_\star=w^2 C(t,t')+\alpha\left[(\id-G)^{-1}(\alpha\sigma^2 E+C)(\id-G^T)^{-1}\right](t,t')\label{eq:cov}
\ee

The matrices $C$ and $G$ in (\ref{eq:kernel}) and (\ref{eq:cov}) are the correlation and response functions of the system, respectively, and are to be determined self-consistently as
\be
C(t,t')=\avg{x(t)x(t')}_\star, ~~ G(t,t')=-\avg{\frac{\delta x(t)}{\delta \nu(t')}}_\star,
\ee
where $\avg{\cdot}_\star$ denotes an average over trajectories of the effective stochastic process (\ref{eq:effproc}), i.e. over realisations of the noise $\{\eta(t)\}$. $E$ is the matrix with all entries equal to one, $E(t,t')=1$ for all $t,t'$.

The analysis proceeds by making a fixed point ansatz $x(t)=x, \eta(t)=\eta, \nu(t)=\nu$ in the effective process, leading to
\be
C(t,t')\equiv Q.
\ee
Furthermore we assume time-translation invariance of the response, i.e. $G(t,t')=G(t-t')$ and define the integrated response as
\be
\chi=\int dt G(t),
\ee
which we require to be finite for the further analysis. This restricts the ansatz to the ergodic
regime of the system, i.e. to model parameters for which
the assumed fixed-point is independent of initial conditions. The
following self-consistent equations of the persistent order parameters
$\{Q,\chi,\nu\}$ can then be derived along the lines of \cite{OD,
galla, gallahebb}:
\BE
\frac{\alpha}{q\sqrt{\lambda}}\left(2u+w^2\Gamma \chi+\frac{\alpha}{1-\chi}\right)&=&\int_{-\infty}^{\Delta} Dz (\Delta-z),\label{eq:firsteq}\\
\frac{Q}{\lambda}\left(2u+w^2\Gamma \chi+\frac{\alpha}{1-\chi}\right)^2&=&\int_{-\infty}^{\Delta} Dz (\Delta-z)^2,\label{eq:secondeq}\\
-\left(2u+w^2\Gamma \chi+\frac{\alpha}{1-\chi}\right)\chi&=&\int_{-\infty}^\Delta Dz.\label{eq:thirdeq}
\EE
Here $Dz=\frac{1}{\sqrt{2\pi}}e^{-z^2/2}dz$ denotes the standard Gaussian measure, and one has $\lambda=w^2 Q+\alpha(\alpha\sigma^2+Q)/(1-\chi)^2$ and $\Delta=\nu/\sqrt{\lambda}$. We note that $\phi=\int_{-\infty}^\Delta Dz=\frac{1}{2}\left(1+\erf\left(\Delta/\sqrt{2}\right)\right)$ describes the fraction of surviving species.

\subsection{Key observables}
We will in the following study the behaviour of the system as a
function of the different model parameters and in particular focus on
the effects of the different components in the setup of the
ecosystem. The above theory allows us to compute the behaviour of the
model in the stable fixed-point regime exactly in the thermodynamic
limit, and to carry out a linear stability analysis to identify the
onset of instability as described below. Theoretical results will be
compared to observations in computer experiments based on a numerical
integration of the replicator equations (\ref{eq:repl}). We here use
the scheme of \cite{OD2}, effectively corresponding to a first order
integrator with dynamical time step. In addition to the above
mentioned fraction of surviving species $\phi$, we will study the
diversity index $D=\frac{\alpha^2}{(q^2Q)}$, closely related to what is
known as Simpson's diversity index in ecology \cite{simpson}. Note
that if the species concentrations were normalised to one the sum
$\sum_i x_i^2$ (i.e. the analogue of $Q=N^{-1}\sum_i x_i^2$) would indicate the
probability that two randomly chosen individuals belong to the same
species. We will also focus on the effectiveness of the use of
resources. To this end one defines
\be\label{eq:hdef}
H=\frac{1}{N^2}\sum_\mu \left(\avg{Q^\mu(t)}_t\right)^2
\ee
with $\avg{\cdot}_t$ a time average in the stationary state. Note that
$\avg {Q^\mu(t)}={\cal O}(N^{1/2})$, so that $H$ is of order one in the
thermodynamic limit. $H$ denotes the efficiency with which the
species make use of the resources present in the system. If $H=0$ then
$\avg{Q^\mu(t)}_t=0$ for all $\mu$, i.e. all resources are optimally
exploited. If however $H>0$, then the use of a fraction of resources
(those with $\avg{Q^\mu(t)}_t\neq 0$) is unbalanced. Our analytical theory allows
us to compute $H$ from the saddle-point equations, and one finds
\be\label{eq:htheory}
H=\alpha\frac{\alpha\sigma^2+Q}{2(1-\chi)^2}
\ee
as in \cite{andreamatteo}.

\subsection{Stability analysis and phase transitions}
The above ansatz of a stable ergodic regime breaks down, when either
fixed points become numerous or suppressed in the thermodynamic
limit. In the first case the system has a large number of (possibly
marginally stable) attractors, and initial conditions determine which
of these is realised. Hence ergodicity is broken. In the second case
the system would not evolve into any fixed point at all at long times.

The breakdown of the fixed point regime can be identified by means of
linear stability analysis on the level of the effective
process. Details of similar calculations can be found in
\cite{OD,galla}. For the present model one finds that system runs into
a unique stable fixed point if
\be\label{eq:instab}
w^2\chi^2+\alpha\frac{\chi^2}{(1-\chi)^2}<\phi,
\ee
and that it becomes unstable when this condition is violated.

Our above fixed-point ansatz also implies the assumption that the
integrated response $\chi$ be finite. A singularity in $\chi$ would
hence signal the breakdown of the ergodic theory and the onset of
memory effects, in the sense that perturbations in the stationary
state do not decay, but remain permanent \cite{Book2}. Simultaneously,
a divergence of $\chi$ necessarily implies $H=0$ (see
eq. (\ref{eq:htheory})), and hence a transition to a phase in which
resources are optimally exploited. Since the right-hand-side $\phi$
(the fraction of surviving species) of Eq. (\ref{eq:thirdeq}) is
bounded ($\phi\in[0,1]$), we find that a divergence of $\chi$ can
occur only if $u=0$ and $w=0$. Thus we expect no phase with optimal
resource exploitation whenever co-operation pressure or direct species
interaction are present. Finally we note that (\ref{eq:thirdeq})
implies $\phi=\alpha$ if $|\chi|\to\infty$ in a model system with
$u=w=0$. Thus, the instability condition (\ref{eq:instab}) is violated
whenever $\chi$ diverges.
\section{Effects of resource variability}\label{sec:var}

We now first examine the effects of the variability of the
resources. To this end, we set the strength of the direct species
interaction $w$ and the co-operation pressure $u$ to zero in this
section, and focus on the behaviour of the model as a function of
$\sigma$. This control parameter $\sigma$ measures the fluctuations of
$Q_0^\mu$ (see Eq. (\ref{eq:defqnull})), i.e. the degree to which the
different resources $\mu=1,\dots,\alpha N$ vary in their bare
abundances $Q_0^\mu$ in the absence of species. For simplicity we keep
$q=1$ throughout this section. This system is the model studied in
\cite{andreamatteo} by static methods. A phase transition was found,
marked by a divergence of the static susceptibility in a replica
symmetric ansatz. We here reproduce this transition from a dynamical
calculation, and present the results of this section mainly for
completeness and to set the scene for the subsequent parts of the
paper. 

\begin{figure}[t!!!]
  \vspace*{10mm} ~~~~~~~~~~~~~~~~\epsfxsize=60mm
  \epsffile{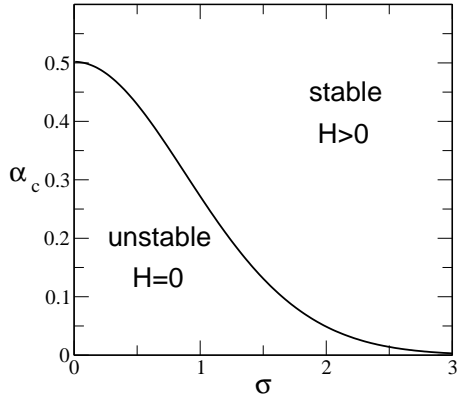} \vspace*{2mm}
  \caption{Phase diagram for the model without direct species interaction and in the absence of co-operation pressure in the $(\alpha,\sigma)$ plane ($q=1$). The integrated response diverges at the phase transition line, and the system becomes sensitive to initial conditions in the unstable phase.}
\label{fig:pg_alpha_sigma}
\end{figure}

\begin{figure}[ht!!!!!]
  \vspace*{10mm} ~~~~~~~~~~~~~~~~\epsfxsize=60mm
  \epsffile{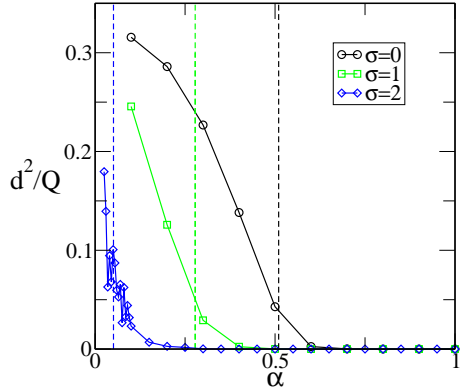} \vspace*{2mm}
  \caption{(Colour on-line) Reduced distance $d^2/Q$ versus $\alpha$ for the model without direct
 species interaction and co-operation pressure ($w=0,u=0$,
 $q=1$). Symbols show results from simulations for $N>200$ species, run
 for $>10000$ discretisation steps and averaged
 over at least $20$ samples of random resource consumption
 $\{\xi_{i}\}$, vertical dashed lines mark the location of the phase
 transition as predicted by the theory.}
\label{fig:distance}
\end{figure}

\begin{figure}[ht!!!]
  \vspace*{10mm} ~~~~~~~~~~~~~~~~\epsfxsize=60mm \epsffile{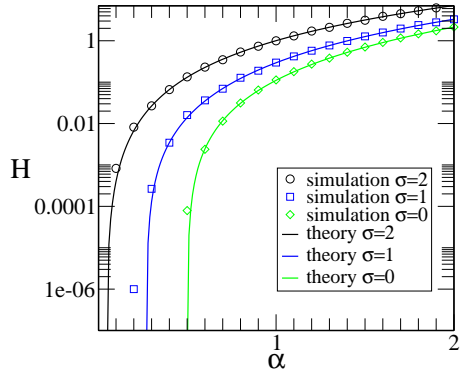}
  \vspace*{2mm} \caption{(Colour on-line) $H$ versus $\alpha$ for the
  model without direct species interaction ($w=0, q=1$). Curves are for
  $\sigma=2,1,0$ from top to bottom. The solid lines are from the
  theory, symbols from simulations ($N=300$, $50$ samples, run for $20000$ steps). $H$ vanishes below $\alpha_c$.}
\label{fig:h_am}
\end{figure}

Fig. \ref{fig:pg_alpha_sigma} shows the phase diagram obtained by
solving the three equations
(\ref{eq:firsteq},\ref{eq:secondeq},\ref{eq:thirdeq}). The transition
point $\alpha_c=\alpha_c(\sigma)$ is identified as the point where the
integrated response $\chi$ diverges. To obtain an interpretation of
this transition in terms of the ergodicity properties of the system,
we run two copies $\{{\bf x(t)}\}$ and $\{{\bf x'(t)}\}$ of the system
with the same realisation of the disorder, but started from different
random initial conditions and measure the distance
$d^2=\avg{N^{-1}\sum_i (x_i(t)-x'_i(t))^2}_t$ between two stationary
states of the system. Thus if $d^2=0$ initial conditions play no role,
while for $d^2>0$ the system is sensitive to the starting
point. Although numerical measurements of $d^2$ can exhibit
finite-size effects, simulations shown in Fig. \ref{fig:distance} are
consistent with an ergodic phase above $\alpha_c$, and with a phase in
which the system is sensitive to initial conditions below $\alpha_c$.
In this second phase the system is still found to evolve into a fixed
point, but stationary points of the dynamics become numerous, and
which one of these is reached asymptotically is determined by initial
conditions, similar to the behaviour of other replicator systems
\cite{DO,OD,OD2}. Fig. \ref{fig:h_am} shows that this ergodic non-ergodic
transition coincides with a transition between a resource-efficient
phase at $\alpha<\alpha_c$ ($H=0$) and an inefficient phase ($H>0$) in
the phase at $\alpha>\alpha_c$.

\begin{figure}[ht!!!]
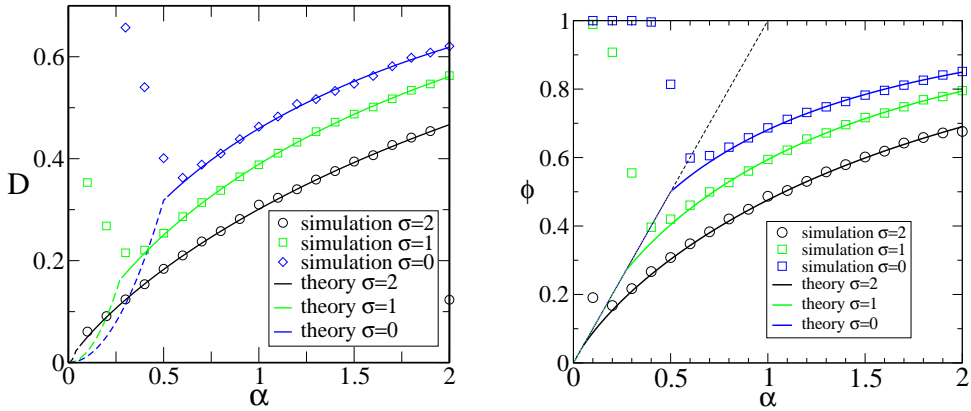

  \vspace*{10mm} \epsfxsize=60mm \epsffile{fig5a.eps} ~~~~
  \epsfxsize=60mm \epsffile{fig5b.eps} \vspace*{2mm}
  \caption{(Colour on-line) Diversity parameter $D=\alpha^2/(Qq^2)$
  (left) and fraction of surviving species $\phi$ (right) versus
  $\alpha$ for the model without inter-species interaction ($w=0,
  q=1$). Curves are for $\sigma=2,1,0$ from bottom to top. The solid
  lines are from the theory, continued as dashed lines into the
  unstable phase in the left panel. Dashed line in right panel marks
  $\phi=\alpha$. Symbols from simulations (parameters as in
  Fig. \ref{fig:h_am}). }
\label{fig:d_am}
\end{figure}

In Fig. \ref{fig:d_am} we report on the diversity of the eco-system as
a function of the resource variability. One finds that the diversity
of the ecosystem is large at a large number of resources per species,
and that the ecosystem becomes less diverse as the number of resources
is reduced. The figures also confirm that the behaviour of
$D=\alpha^2/(q^2Q)$ is similar to the one of the fraction of surviving
species $\phi$, hence verifying the role of $D$ as a measure of the
diversity of the ecosystem. In the following sections we will hence
focus on $\phi$.  As anticipated in the introduction, results do not depend on the specific shape of the distribution of the $\{\xi_i^\mu\}$, as only their first and second moments enter in the derivation of the effective dynamics. We have explicitly confirmed this in simulations, which show that measurements of $H$ and $D$ of systems in which the $\{\xi_i^\mu\}$ follow flat, exponential, bimodal and power-law distributions with suitable first and second moments fall precisely on the lines obtained from the theory in Figs \ref{fig:h_am} and \ref{fig:d_am}.

The left panel of Fig. \ref{fig:d_am} furthermore confirms that
$\phi=\alpha$ at the transition with diverging susceptibility
$\chi$. Similar transitions in static contexts can be identified by
the divergence of the static susceptibility in a replica symmetric
approach \cite{andreamatteo}. The occurrence of this transition has a
geometrical interpretation similar to what is know in the context for example of Minority Games \cite{continuum, desanctisgalla, Book1,Book2}. In the
absence of co-operation pressure and direct species interaction, the
fitness $f_i$ in expression (\ref{eq:repl}) is of the form $f_i[{\bf
Q}]=N^{-1}\sum_\mu Q^\mu(t)\xi_i^\mu$, i.e. a linear combination of
the $P$ $N$-dimensional vectors
${\mathbf{\xi}^\mu}=(\xi_1^\mu,\cdots,\xi_N^\mu)$. The dynamics of the
system thus can only wash out perturbations within the space spanned
by the $\alpha N$ vectors ${\bf
\xi^\mu}$,$\mu=1,\dots,\alpha N$. Disregarding the $(1-\phi)N$ extinct species, the underlying dynamical system has $\phi N$ effective degrees of freedom. Extinct species are typically stably extinct with respect to perturbations, see also \cite{OD}. The space of all potential external perturbations is hence $\phi N$-dimensional. Thus if $\alpha<\phi$ some of those perturbations can not be removed by the dynamics, and ergodicity breaking occurs. 

The existence of a phase with $H=0$ at $\alpha<\alpha_c$ can be
interpreted similarly. In the absence of direct interaction and
co-operation pressure one has ${\cal H}$ and $H$ coincide up to
pre-factors, and the dynamics minimises this Lyapunov
function. Attaining the absolute minimum $H=0$ implies
$\avg{Q^\mu(t)}_t=0$ for all $\mu$ via (\ref{eq:hdef}). This
constitutes a system of $\alpha N$ constraints. With $\phi N$
effective degrees of freedom available, these conditions can be met if
$\alpha<\phi$, but not above the transition point defined by
$\phi(\alpha_c)=\alpha_c$.

\section{Effects of direct species interaction}\label{sec:J}

Animals have not only the resource competition but also have the 
direct species interaction, like prey-predator, co-operation,
competition and so on.
 In this section we study the effects of direct interaction between
species, as controlled by the model parameter $w$. In order to focus
on the impact of this model parameter we set $u=0$ throughout this
section. We also limit the discussion to the case $q=1$.

\begin{figure}[t!!!]
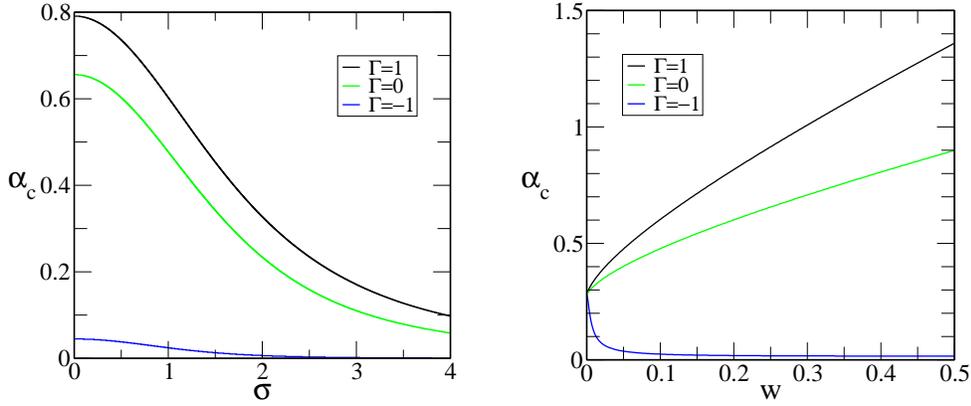

  \vspace*{10mm}
\epsfxsize=60mm \epsffile{fig6a.eps} 
~~~~~\epsfxsize=60mm \epsffile{fig6b.eps}
\vspace*{2mm}
  \caption{(Colour on-line) {\bf Effect of direct species
  interactions.} {\bf Left:} Phase diagram in the $(\sigma, \alpha)$
  plane ($w=0.1$, $q=1$, $u=0$). The curves are obtained from Eq. (\ref{eq:instab}) and are shown for $\Gamma=1,0,-1$ from top to bottom. System is stable above the respective curves, and unstable below. {\bf Right:} Phase diagram in the
  $(w,\alpha)$ plane. $\sigma=1$, $q=1$ and $u=0$. Curves are for
  $\Gamma=1,0,-1$ from top to bottom. System is stable above the respective curves. }
\label{fig:pg_alpha_sigma_J}
\end{figure}

In Fig. \ref{fig:pg_alpha_sigma_J} we depict the phase behaviour of
the system as direct species interaction is introduced. As predicted
by the theory we find that the integrated response $\chi$ is finite in
case $w>0$ for all tested values of the model parameters, the
transition lines in Fig. \ref{fig:pg_alpha_sigma_J} hence mark an
instability at which $H$ remains positive. They are obtained from
Eq. (\ref{eq:instab}). The figure demonstrates that the phase diagram
is indeed affected by the direct interaction between species and by
the symmetry of the couplings $\{w_{ij}\}$.  The left panel shows that
even a relatively moderate direct interaction of strength $w=0.1$ can
have a significant effect: symmetric ($\Gamma=1$) and asymmetric
($\Gamma=0$) interaction reduce the stable area while antisymmetric
($\Gamma=-1$) interaction expands the stable region, compared with the
case without direct interaction ($w=0$) shown in
Fig. \ref{fig:pg_alpha_sigma}.  This is confirmed in the right panel
of Fig. \ref{fig:pg_alpha_sigma_J}: For symmetric and uncorrelated
interaction $\alpha_c$ increases with increasing $w$ so that direct
interaction tends to make the system increasingly less stable. For
negatively correlated interaction ($\Gamma=-1$) on the other hand,
$\alpha_c$ is a decreasing function of $w$, indicating that
prey-predator-type interactions stabilise the ecosystem. One might
speculate that for that reason, food-webs with this type of
interaction may be more likely to be observed in nature than
others. For $\Gamma=-1$ we also find $\alpha_c$ approaches zero for
large values of $w$ indicating that there is no unstable region in the
limit of $w\to\infty$, which is consistent with marginally stable
dynamics in the antisymmetric random replicator model without resource
competition
\cite{Chawanya_Tokita_2002}.

\begin{figure}[ht!!!]
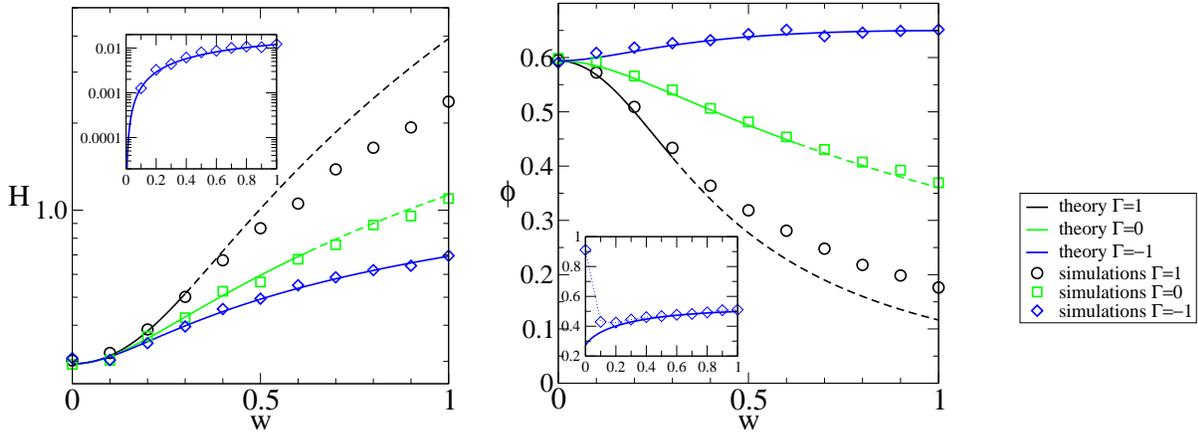

	\vspace{10mm} \epsfxsize=60mm\epsffile{fig7a.eps}
	~~~~~\epsfxsize=93mm \epsffile{fig7b.eps} \vspace*{2mm}
	\caption{{\bf Effect of direct species interactions:} efficiency of
	resource exploitation $H$ (left) and fraction of surviving
	species $\phi$ (right) versus $w$ ($u=0$,
	$\alpha=q=\sigma=1$). Solid lines are from the theory in the
	stable phase, continued as dashed lines into the unstable
	phases. Symbols are from simulations, circles, squares and
	diamonds are $\Gamma=1,0,-1$ respectively ($N>200$, $>20$ samples, $>10000$ iterations). The insets show the case $\alpha=0.2,\Gamma=-1$ for comparison. }
\label{fig:H_phi_w}
\end{figure}

The left panel of Fig. \ref{fig:H_phi_w} shows that the efficiency of
resource exploitation is reduced as direct species-interaction is
introduced, and is consistent with the predicted absence of a phase in
which $H=0$. The effect is stronger for correlated couplings than for
negatively correlated ones. The effects of the direct couplings on the
diversity of the eco-system is shown in the right panel of
Fig. \ref{fig:H_phi_w}. One observes relatively little effect for the
case of antisymmetric couplings, but a strong reduction of diversity as
uncorrelated or positively correlated couplings are
introduced. Crucially we here find that $H$ and $\phi$ are smooth
functions of $w$ as long as $\alpha>\alpha_c(w=0)$. In particular no
singularities are observed as $w\to 0$. This is different in the case
$\alpha=0.2<\alpha_c(w=0)\approx 0.27$, as shown in the insets of Fig.
\ref{fig:H_phi_w}. Here $H\to 0$ as $w\to 0$ and the integrated response
diverges. Simulations at finite $N$ reveal non-monotonous behaviour of
$\phi$ at $w=0^+$. While we cannot fully rule out finite-size effects
similar discontinuities of order parameters have been found in the
context of so-called grand canonical Minority Games
\cite{Book1,Book2}. The apparent discontinuity of $\phi$ will become
even more pronounced in the context of co-operation pressure, as
discussed below.

\section{Effects of co-operation pressure}\label{sec:u}
We now turn to the effects of the co-operation pressure $u$ on the
behaviour of the model. 
We again limit the discussion to the case
$\sigma=q=1$, and consider the system both with and without direct
species interaction.

\subsection{No direct species interaction}
\begin{figure}[t!!!]
  \vspace*{10mm} ~~~~~~~~~~~~~~~~\epsfxsize=60mm
  \epsffile{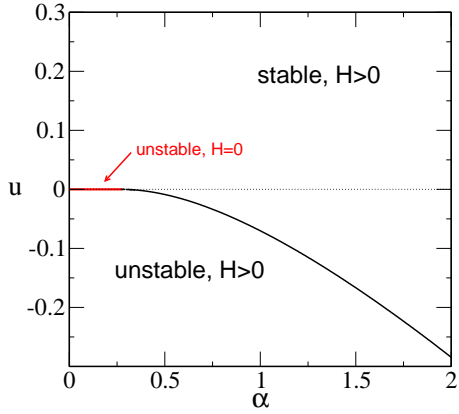} \vspace*{2mm}
  \caption{Phase diagram for the model with co-operation pressure ($w=0$, $q=\sigma=1$). 
 }
\label{fig:phasediagram_u_w0}
\end{figure}

\begin{figure}[t!!!]
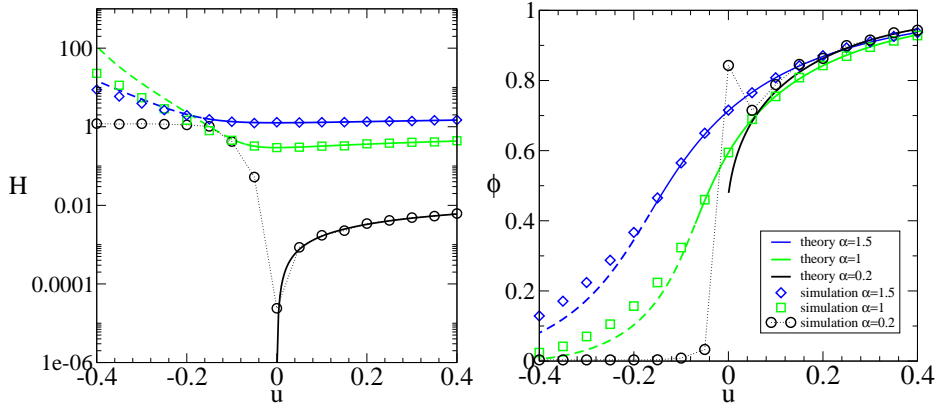

	\vspace{10mm}
  \epsfxsize=62mm\epsffile{fig9a.eps}
  \epsfxsize=60mm \epsffile{fig9b.eps} \vspace*{2mm} 
  \caption{(Colour on-line) {\bf Effects of co-operation pressure:} Efficiency of resource exploitation $H$ (left) and fraction of surviving species $\phi$ (right) versus co-operation pressure $u$ ($w=0$, $q=\sigma=1$). Solid lines are from the theory in the stable phase, for $\alpha=1,1.5$ continued as dashed lines into the unstable phases with finite integrated response. Symbols are from simulations (circles correspond to $\alpha=0.2$, squares to $\alpha=1$, diamonds to $\alpha=1.5$) with $N=300$, run for $20000$ iteration steps, averaged over $50$ samples. Markers for $\alpha=0.2$ have been connected as a guide to the eye.}
\label{fig:H_u_w0_2}
\end{figure}

The phase behaviour of the system at $w=0$ is depicted as a function
of the co-operation pressure in Fig. \ref{fig:phasediagram_u_w0}. For
completeness we extend the discussion to positive and negative values
of the co-operation pressure $u$, although only $u>0$ carries specific
ecological meaning (plants however can grow without predation, which
might be modelled by a positive self-interaction, corresponding to a
negative co-operation pressure in the present model). Most interestingly the
phase with optimal exploitation of resources is limited to the
interval $u=0,\alpha\in[0,0.27]$ on the $u=0$ axis. In particular, as
mentioned above, any positive or negative amount of co-operation
pressure removes the phase with $H=0$.  Furthermore as observed in
Fig. \ref{fig:phasediagram_u_w0}, the eco-system is fully stable at
all $\alpha$ for any positive co-operation pressure, even for
infinitesimally small $u>0$. For $\alpha\gtrsim0.27$ an
unstable phase can only be found at $u<u_c(\alpha)<0$.
Fig. \ref{fig:H_u_w0_2} confirms that $H>0$ throughout this phase. As
expected $\phi$ grows monotonically with $u$, the co-operation
pressure $u$ acts as a force driving the system into the interior of
the simplex (\ref{eq:norm}). For low or negative values of $u$
on the other hand the fraction of surviving species is low. Our simulations seem
to indicate that $\phi$ is continuous as $u\downarrow 0$ for
$\alpha>\alpha_c(u=0)$, but that a discontinuity may be present at
lower values of $\alpha$. This is similar to the behaviour of $\phi$
at low $\alpha$ as $w\downarrow 0$ discussed above (see inset of
Fig. \ref{fig:H_phi_w}). As shown in the right panel of
Fig. \ref{fig:H_u_w0_2} $\phi$ attains values close to zero for
$\alpha=0.2$ and $u<0$, whereas the fraction of surviving species is
clearly positive at positive $u$. While our simulations are
potentially prone to finite-size effects, the data presented is
consistent with a first order phase transition. Simulations
furthermore indicate that $\phi$ might actually vanish at small enough
$\alpha$ and negative co-operation pressure indicating the possible
existence of a phase in which only a sub-extensive number of species
survives. Such behaviour has previously been reported for the case of
higher-order interaction in
\cite{galla}. Due to the limited relevance of negative co-operation
pressure we have however not conducted a more detailed analysis of
these observations, and can at this stage not fully confirm the
existence of such a phase in this system of two-body interaction.

\subsection{With direct species interaction}

\begin{figure}[t!!!]
  \vspace*{10mm} ~~~~~~~~~~~~~~~~\epsfxsize=60mm
  \epsffile{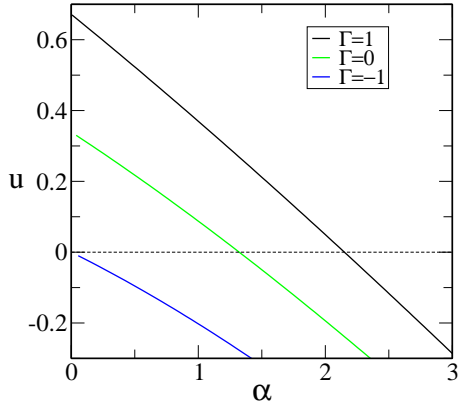} \vspace*{2mm}
  \caption{(Colour on-line) Phase diagram for the model with species interaction ($w=1$)
 and co-operation pressure. Resource variation is set to $\sigma=1$. Curves
 show the onset of instability for $\Gamma=1,0,-1$ from top to bottom,
 with stable phases to the top-right, unstable ones to the lower
 left. }
\label{fig:phasediagram_u_J}
\end{figure}

\begin{figure}[t!!!]
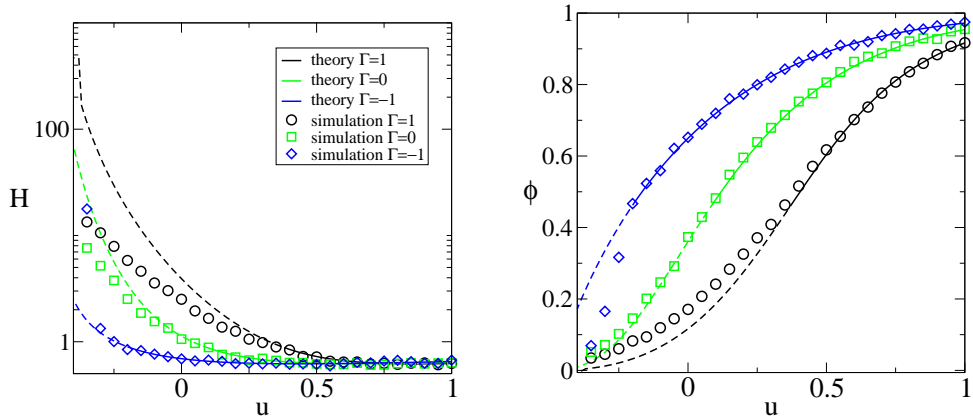

	\vspace{10mm} \epsfxsize=60mm\epsffile{fig11a.eps}
	~~~~~\epsfxsize=60mm \epsffile{fig11b.eps} \vspace*{2mm}
	\caption{(Colour on-line) {\bf Effects of co-operation
	pressure:} Efficiency of resource exploitation $H$ (left) and
	fraction of surviving species $\phi$ (right) versus
	co-operation pressure $u$ for model with direct species
	interaction ($w=\alpha=q=\sigma=1$). Solid lines are from the
	theory in the stable phase, continued as dashed lines into the
	unstable phases. Symbols are from simulations, circles
	correspond to $\Gamma=1$, squares to $\Gamma=0$, diamonds to
	$\Gamma=-1$ (simulations are performed for $N=200$, run for $10000$ iteration steps, averaged over $20$ samples of the disorder). }
\label{fig:H_u_with_j}
\end{figure}

The phase structure of the model with co-operation pressure and direct
species interaction is shown in
Fig. \ref{fig:phasediagram_u_J}. Stable phases are found at large
positive co-operation pressures and large relative numbers of
resources, and either a reduction of $u$ or $\alpha$ can induce
instability. In line with earlier observations anti-symmetry in the direct
species interactions tends to stabilise the eco-system, at full
anti-correlation a stable fixed-point regime is found for any $u>0$ at
any $\alpha$, whereas unstable regimes can be found for $\Gamma>-1$
even at positive co-operation pressure. The left panel of
Fig. \ref{fig:H_u_with_j} finally shows that $H$ remains positive
throughout all tested parameter ranges if $w>0$. The right panel
demonstrates that again $\phi$ is an increasing function of the
co-operation pressure $u$. In contrast with the system at $w=0$ no
discontinuities in the order parameters are observed, as the
transition with diverging integrated response is absent.

\section{Effects of dilution}\label{sec:dil}
Animals, of course,  do not have the all-to-all interaction.
We now turn to a discussion of the effects of diluting the interaction
web between species. We restrict the discussion to the case without
direct species interaction and without co-operation pressure, i.e. we
consider only $u=w=0$. We furthermore follow the philosophy of
introducing dilution in the context of neural networks \cite{Cool00b}
and in random replicator models \cite{gallahebb}, and assume that only
a fraction $c\in(0,1]$ of interaction links between species is
present. If an interaction between species $i$ and $j$ is present,
then we take it to be determined by their respective use of resources,
following \cite{andreamatteo}. More specifically, we write

\be
\frac{\dot x_i(t)}{x_i(t)}=\frac{1}{N}\sum_{\mu=1}^{\alpha c N} \xi_i^\mu Q_i^{\mu}(t)+\nu(t)
\ee
where
\be
Q_i^\mu(t)=Q_0^\mu-\sum_{j=1}^N\frac{c_{ij}}{c}\xi_j^\mu x_j(t)
\ee
is the amount of resource $\mu$ available to species $i$ at time
$t$. The coefficients $c_{ij}$ denote the dilution of the interactions
and take values $0$ and $1$, indicating a particular link to be absent
or present, respectively. An absent link could for example correspond
to a geographic separation between species. The $c_{ij}$ are here
taken to be random, and we choose any $c_{ij}$ to be equal to one with
probability $c$, and equal to $0$ with probability $1-c$. Consequently
we have
\be
c=\avg{c_{ij}}_c=\avg{c_{ji}}_c
\ee
for any pair $i<j$, where $\avg{\dots}_c$ denotes an average over realisations of the dilution. Note that $\avg{c_{ij}^2}_c=c$. Correlations in the interaction network are then introduced by the requirement that
\be
\avg{c_{ij}c_{ji}}_c-c^2=\gamma c(1-c)
\ee
with $\gamma\in[0,1]$. $\gamma=1$ corresponds to an undirected symmetric network of interactions with $c_{ij}=c_{ji}$ for all $i<j$. For $\gamma=0$ $c_{ij}$ and $c_{ji}$ are uncorrelated, and the links in the interaction web are hence directed ones. Ecologically realistic cases presumably correspond to $\gamma\approx 1$, for completeness we extend the statistical mechanics analysis of the dilute model to general values $\gamma\in[0,1]$. Finally, we note that following the conventions in the literature we write the number of resources as $P=\alpha c N$ in this section, and that we take self-interactions to be present for all species, i.e. we have $c_{ii}=1$ for all $i=1,\dots,N$. We also note that we use $Q_0^\mu=\sigma\sqrt{P}\zeta^\mu$ and $\overline{\xi_i^\mu}=0$ along with the normalisation $N^{-1}\sum_i x_i(t)=\alpha c$  in this section \footnote{The modification to the statistics of the $\{Q_0^\mu,\xi_i^\mu\}$ is necessary to guarantee a well defined thermodynamic limit. While in the fully connected model all terms of order higher than $N^0$ drop out in the dynamical action due to the overall normalisation of species concentrations, this is would no longer be the case in the dilute model. If the statistics of the $\{Q_0^\mu,\xi_i^\mu\}$ were not modified, $N$ different normalisation constraints would be required, due to different local interaction `neighbourhoods' of species. The model specifications used in this section make sure that such terms do not appear.}.

\begin{figure}[t!!!]
  \vspace*{10mm} ~~~~~~~~~~~~~~~~\epsfxsize=60mm
  \epsffile{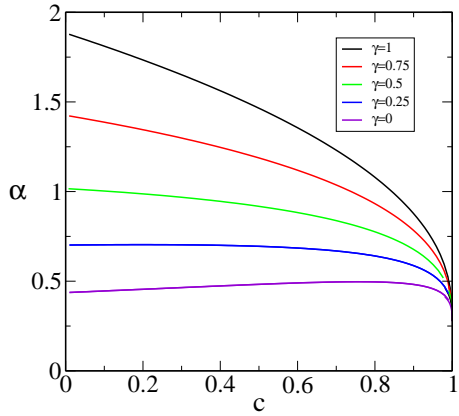} \vspace*{2mm}
  \caption{(Colour on-line) Phase diagram for the dilute model ($\sigma=1$). The curves show the transition lines below which fixed points become unstable. $\gamma=1,0.75,0.5,0.25,0$ from top to bottom. No divergence of the susceptibility $\chi$ is observed for $c<1$. At $c=1$ one reproduces the transition of the model of \cite{andreamatteo}, $\alpha_c(c=1)\approx 0.27$. }
\label{fig:pg_dilute}
\end{figure}

\begin{figure}[t!!!]
  \vspace*{10mm} ~~~~~~~~~~~~~~~~\epsfxsize=60mm \epsffile{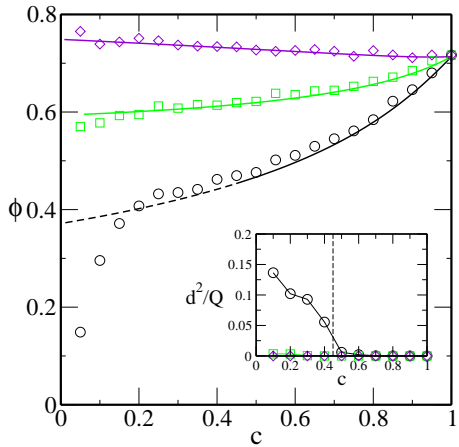}
  \vspace*{2mm} \caption{(Colour on-line) {\bf Effects of dilution:}
  Fraction of surviving species $\phi$ versus $c$ for dilute model at
  $\alpha=1.5$ ($\sigma=1$). $\gamma=0,0.5,1$ from top to
  bottom. Solid lines are from theory in the stable phase, and have
  been continued as dashed lines into the unstable phase (where the
  theory can no longer be expected to be accurate). Symbols are from
  simulations ($N=300$, run for $40000$ time steps, averages over at
  least $10$ samples are taken; for small values of $c$ simulations
  may exhibit finite-size effects; also equilibration effects and
  sample to sample fluctuations cannot fully be excluded). The inset
  shows $d^2/Q$ versus $c$ obtained from simulations for the same
  model parameters, and demonstrates the presence of a transition for
  $\gamma=1$ (circles). For $\gamma=0,0.5$ $d^2/Q\approx 0$ in line
  with the theory which for those values of the symmetry parameter
  predicts the system to be stable for all $c$.}
\label{fig:phi_dilute_a_1.5}
\end{figure}

The analysis of the dilute model is straightforward and can be performed along the lines of \cite{Cool00b,Verbeiren,gallahebb}. The effective process reads:
\be
\hspace{-2.5cm}\dot x(t)=x(t)\left(-\alpha (1-c)x(t')-\alpha\int_{t_0}^t dt' \left[c (\id-G)^{-1}+\gamma(1-c)G\right](t,t')x(t')-\eta(t)+\nu(t)\right)
\ee
where
\be
\avg{\eta(t)\eta(t')}=\alpha\left[c(\id-G)^{-1}(\alpha c\sigma^2 E+C)(\id-G^T)^{-1}+(1-c)C\right](t,t')
\ee
and with all other definitions as in the fully connected model.

The resulting phase diagram is depicted in Fig. \ref{fig:pg_dilute}. As
shown, correlated dilution (roughly $\gamma\ge0.5$) increases the
numerical value of $\alpha_c$, and hence reduces the stable regime of
the system compared with the fully connected model. At largely
uncorrelated dilution $\gamma\le 0.5$ the location of the phase
transition $\alpha_c$ shows only a weak dependence on the degree of
dilution $c$. This behaviour is also reflected in
Fig. \ref{fig:phi_dilute_a_1.5}, where we focus on the system at
$\alpha=1.5$ and depict the fraction of surviving species as a
function of the connectivity $c$ at different values of the symmetry
parameter $\gamma$. For $\gamma$ smaller than roughly one half,
diluting the network of species does not seem to affect the stationary
state significantly. The phase transition is absent, and the system
always reaches a unique stable fixed point at this value of $\alpha$,
irrespectively of $c$.  Uncorrelated dilution furthermore has only
little effect on the diversity whereas as highly correlated
interaction network can affect the ecosystem significantly, and
reduces the number of survivors. This is in-line with our earlier
observations on the effect of direct species interaction at different
degrees of symmetry, see Fig. \ref{fig:H_u_with_j}.

\section{Concluding remarks and outlook}
In summary we have used tools from disordered systems theory to study
a stylised model of a simple eco-system, composed of a set of species
competing for an amount of limited resources, and which at the same
time are subject to direct inter-species competition.

The dynamical
system of corresponding replicator equations has been addressed by
path integral techniques, allowing us in particular to study cases of
asymmetric interaction between species (corresponding to prey-predator
relations), where there is no Lyapunov function governing the
dynamics, and where static approaches are hence inapplicable. 

We find that this simple model eco-system displays a rich spectrum of
features, and interesting phase behaviour separating stable from
unstable regimes. Our main findings can be summarised as follows: (i)
in absence of direct species interaction and co-operation pressure the
fully connected model displays a transition between a phase in which
initial conditions are irrelevant and a non-ergodic phase. This
transition is also marked by a change of the efficiency of resource
exploitation. In the unstable phase resources are used optimally,
while this is not the case in the stable phase. (ii) The introduction
of either direct species interaction, co-operation pressure or
dilution alters the type of transition observed, in particular the
fully efficient phase is removed. One still finds a phase boundary
separating a stable ergodic fixed point regime from a non-ergodic
phase. (iii) For symmetric couplings the non-ergodic phase is marked
by an exponential number of marginally stable fixed points, see also
\cite{OD,OD2}, and initial conditions determine which of these is
reached in the long run leading to the observed ergodicity
breaking. At asymmetric or partially asymmetric coupling (induced by
either direct interaction or dilution) no fixed point is reached in
the unstable phase. Instead the trajectories of the system remain
volatile and potentially chaotic. (iv) We observe a general tendency
of increased stability when asymmetric (or anti-symmetric) interaction
is introduced. This is the case both for direct species interaction
and dilution. While the range of stability is then increased, no
significant effects on the diversity of the eco-system are found. (v)
The introduction of symmetric interaction or dilution can reduce the
stability of the system significantly and at the same time also lead
to a reduced diversity of its stationary population structure. (vi) In
the absence of direct species interaction the effects of co-operation
pressure can be drastic, and in particular the system is stable at any
even infinitesimal amount of co-operation pressure. At small
(relative) numbers of resources over species order parameters can
display discontinuities as the co-operation pressure tends to zero.

The model studied in the present paper and its phase behaviour are
furthermore interesting from the statistical mechanics point of
view. As detailed above the transition between a resource-efficient
and an inefficient phase in the model without direct species
interaction or co-operation pressure has a geometrical interpretation
previously identified e.g. in the context of the dynamics of the
Minority Game \cite{continuum, desanctisgalla}. This geometrical
picture breaks down as soon as direct interaction or co-operation
pressure are introduced, hence the absence of a transition at
diverging integrated response and of the fully efficient phase. The
present model may hence serve as a starting point for attempts to
fully classify interacting agent models according to the presence or absence
of phases with optimal resource exploitation. A close relation to the
presence or otherwise of replica-symmetry breaking and to the geometry
of the manifold of stationary states is here to be expected.

Extensions of the present model might include adding a third or
further trophic levels, temporally fluctuating resource availability
(e.g. along the lines of \cite{desanctisgalla}) or the introduction of
further heterogeneity of the species. It is likely that this will
alter the phase diagram, and might affect the stability or otherwise
of the eco-system. Furthermore the computation of species abundance
distributions (SAD), as introduced by Fisher et al \cite{Fisher} and by Preston
\cite{Preston} might be an interesting issue for future
work. SAD have been measured and compared to log-normal and log-series
distributions known in ecology for example in the model of
\cite{McKane2000}. The work of \cite{TokitaPRL} demonstrates that
random replicator models can yield SAD similar to left-skewed
log-normal distributions. Given the presence of a phase transition in
the model discussed here it would be particularly interesting to study
{\em finite} systems near the transition, resulting in potential
non-Gaussian features and fat-tailed abundance distributions (see
\cite{Book1,Book2} and references therein for similar critical
fluctuations in Minority Game models near their phase transitions). In
order to address the resulting topological structure and distribution
of coupling strengths it might also be interesting to study the food
web resulting from the present model in more detail. In particular, as
seen above, some species die out asymptotically, inducing a reduced
coupling matrix restricted to survivors only. While species can not
change their foraging strategies and no new links between species can
created in our model, this extinction dynamics might lead to
non-trivial, potentially correlated effective interaction strengths
distributions among survivors at stationarity. In \cite{Tokita2} a
dominance of prey-predator pairs in the set of surviving species has
for example been identified in the context of replicator systems with
quenched Gaussian interactions.

Furthermore, it would be interesting
to extend the analysis of the dilute model to more realistic
finite-connectivity cases on complex networks \cite{Albert}. So far we
have only addressed dilute Erd\"os-Reyni type networks with an {\em
extensive} number of connections per node. Complex networks with
scale-free degree distribution \cite{Albert,Montoya} and the
small-world property \cite{watts} or other
structures might here be of more biological relevance
\cite{McKane2000,sole2002,McKane_Drossel_2005}, in an approach to approximate dynamically
evolved networks by static quenched ones. Extension to such sparse
networks might require to study cases with only a {\em finite} number
of interactions per species. This is challenging as the resulting
effective dynamical theories do not close on the level of two-time
order parameters. Still it would be interesting to examine the effects
of network topology and degree sequence on the stability or otherwise
of the model eco-system, as a first step potentially relying on
numerical simulations or on replica approaches and the cavity-method
\cite{weigt} in order to study the statics of eco-systems with
symmetric couplings.

\section*{Acknowledgements}
The authors would like to acknowledge fruitful discussions with A De
Martino and M Marsili. YY thanks the condensed matter and statistical
physics group at ICTP for hospitality.  This work was supported by EU
NEST No. 516446 COMPLEXMARKETS, by IST STREP GENNETEC, contract number
034952 and by an Osaka University Scholarship (short-term
student dispatch program). YY and KT are partially supported by The
21st Century COE program `Towards a new basic science: depth and
synthesis'.  KT acknowledges support by grants-in-aid from MEXT, Japan
(No. 14740232 and 17540383) and through the priority area `Systems
Genomics'.


\end{document}